\documentclass[11pt]{article}
\usepackage{graphicx,amssymb,amsmath,amsfonts}
\usepackage{epsfig}
\usepackage[margin=3cm]{geometry}
\newdimen\rotdimen
\def\vspec#1{\special{ps:#1}}
\def\rotstart#1{\vspec{gsave currentpoint currentpoint translate
   #1 neg exch neg exch translate}}
\def\rotfinish{\vspec{currentpoint grestore moveto}}
\def\rotr#1{\rotdimen=\ht#1\advance\rotdimen by\dp#1%
   \hbox to\rotdimen{\hskip\ht#1\vbox to\wd#1{\rotstart{90 rotate}%
   \box#1\vss}\hss}\rotfinish}
\def\rotl#1{\rotdimen=\ht#1\advance\rotdimen by\dp#1%
   \hbox to\rotdimen{\vbox to\wd#1{\vskip\wd#1\rotstart{270 rotate}%
   \box#1\vss}\hss}\rotfinish}%
\def\rota#1#2{\rotdimen=\ht#1\advance\rotdimen by\dp#1%
   \hbox to\rotdimen{\hskip\ht#1\vbox to\wd#1{\rotstart{#2 rotate}%
   \box#1\vss}\hss}\rotfinish}
\def\rotu#1{\rotdimen=\ht#1\advance\rotdimen by\dp#1%
   \hbox to\wd#1{\hskip\wd#1\vbox to\rotdimen{\vskip\rotdimen
   \rotstart{-1 dup scale}\box#1\vss}\hss}\rotfinish}%
\def\rotf#1{\hbox to\wd#1{\hskip\wd#1\rotstart{-1 1 scale}%
   \box#1\hss}\rotfinish}%

\renewcommand{\baselinestretch}{1.0}

\begin{document}

\def\Im{\mbox{ Im}}   
\def\Re{\mbox{ Re}}
\def\sqrtm1{\mbox{ i}}   
\def\bigOh{\mbox{${\cal O}$}}
\def\Pen{\mbox{${\it Pe}$}}
\def\Deff{\mbox{$D_{\mbox{eff}}$}}
\def\measd{\mbox{d}}
\def\uv{\nobreak\mbox{${\bf \hat{e}}$}}
\def\su{\nobreak\mbox{{\bf S}}}
\def\lp{\mbox{\Large $($}}
\def\rp{\mbox{\Large $)$}}
\def\etal{\mbox{\it et. al. }}
\def\ie{\mbox{\it i.e. }}
\def\sNm1{\mbox{$\scriptstyle{_{{N\!-\!1}}}$}}
\def\sN{\mbox{$\scriptstyle{_{N}}$}}
\newcommand{\p} {{\rm \partial}}
\def\K{\mbox{K}}
\def\sn{\mbox{sn}}
\def\htc{\mbox{\textit{\textsf{h}}}}


\title{
\vspace{-1in}
Critical thickness of an optimum extended surface
characterized by uniform heat transfer coefficient}
\author{Theodoros Leontiou\footnote{\mbox{E-mail: eng.lt@fit.ac.cy}} ,
        $~$ and Marios M. Fyrillas$^{\ast ,}$\footnote{\mbox{E-mail: marios.fyrillas@nu.edu.kz}}\\
        {\small $^{\ast}$General Department, Frederick University,}
        {\small 1303 Nicosia, Cyprus.}\\
        {\small $^\dag$Department of Mechanical Engineering,}\\
        {\small Nazarbayev University, Astana 010000, Republic of Kazakhstan.}
                                                }
\maketitle
\begin{abstract}

We consider the heat transfer problem associated with
a periodic array of extended surfaces (fins)
subjected to convection heat transfer with
a uniform heat transfer coefficient.
Our analysis differs from the classical
approach as (i) we consider two-dimensional heat conduction
and (ii) the base of the fin is included in the heat transfer
process.
The problem is modeled as an arbitrary two-dimensional channel
whose upper surface is flat and isothermal,
while the lower surface has a periodic array
of extensions/fins which are subjected to heat convection
with a uniform heat transfer coefficient. Using the generalized
Schwarz-Christoffel transformation the domain is mapped
onto a straight channel where the heat conduction problem is
solved using the boundary element method. The boundary element
solution is subsequently used to pose a shape optimization
problem, i.e. an inverse problem, where the objective function
is the normalized Shape Factor and the variables of the optimization
are the parameters of the Schwarz-Christoffel transformation.
Numerical optimization suggests that the optimum fin is infinitely
thin and that there exists a critical Biot number that characterizes
whether the addition of the fin would result in an enhancement
of heat transfer. The existence of a critical Biot number was investigated
for the case of rectangular fins. {\bf It is concluded that
a rectangular fin is effective if its thickness is less than} $1.64~k/\htc$,
where the $\htc$ is the heat transfer coefficient and $k$ is the thermal
conductivity. This result is independent of both the thickness of the
base and the length of the fin.

\end{abstract}

\subsubsection*{Keywords}
Optimum fin design; shape Optimization; Inverse Design; Heat Convection;
uniform heat transfer coefficient; generalized Schwarz-Christoffel transformation;
Laplace equation.

\renewcommand{\baselinestretch}{1.0}

\begin{table}[h]
\begin{tabular}{|ll|} \hline
{\bf Nomenclature}  & \\
                    & \\
$Bi$                & Biot number $Bi= \htc L /k$ \\
$H$                 & dimensionless height of the slab/channel (dimensionless) \\
$H_b$               & thickness of the base (dimensionless) \\
$H_f=H-H_b$         & length of fin/extended surface (dimensionless) \\
$h$                 & height of the slab/channel in the transformed domain (dimensionless) \\
\htc                & heat transfer coefficient ($\mbox{W/(m$^2$ K)}$)\\
$\mbox{i}$          & imaginary unit \\
$k$                 & thermal conductivity ($\mbox{W/(m K)}$) \\
$P$                 & arc-length of the periodic geometry (dimensionless) \\
$L$                 & period of the geometry (m) \\
$T$		            & temperature (dimensionless) \\
$x,y$  	            & coordinates of the physical plane (dimensionless) \\
$z=x+\mbox{i}\,y$   & complex coordinate of the physical plane (dimensionless)\\
$z_i$               & vertices in the physical plane \\
                    & \\
{\em Greek symbols} & \\
$\alpha_{i}$	    & turning angles \\
$w$                 & complex coordinate of the transformed domain (dimensionless)\\
$w_{i}$             & image of $z_i$ vertices in the transformed domain \\
$w_N$               & period of the geometry in the transformed domain \\
                    & \\
{\em Subscripts}    & \\
$i$		            & related to the $i$-th vertex \\
                    & \\
{\em Diacritic}     & \\
$\wedge$	        & the variable is normalized with $w_N$ \\
              \hline
\end{tabular}
\end{table}


\section{Introduction}

Assuming isothermal boundary conditions, any extension
from a surface would result in a reduction in heat transfer
rate. This statement can be deduced from geometrical inclusion
theorems \cite{Lavrentev65}. It implies that the
heat transfer rate across a two-dimensional channel, bounded
by isothermal surfaces, is higher than that across a
similar channel whose surfaces are extended, i.e.
by adding fins \cite{Incropera}, assuming that the temperature difference
between the surfaces
remains the same. Hence, within the approximation of isothermal
conditions and constant temperature difference,
the addition of fins would simply reduce the heat transfer
rate!
Because isothermal conditions
can be realized in the limit of infinite Biot number, the following
question is raised: {\bf When is an extended surface (fin) effective?}
This question will be
elucidated in this paper, where we show that the limit of
infinite Biot number is singular \cite{Fyrillas00,Fyrillas14},
and a critical thickness exists \cite{Fyrillas11}.

This paper considers the fundamental problem of finding the optimum
shape of an extended surface/fin such that the heat transfer rate
is maximized. In particular, we consider the inverse design problem
associated with two-dimensional (2D) heat conduction in a
finite 2D periodic channel/slab with a flat isothermal
upper boundary and a periodic lower boundary which is subjected
to convection with a constant heat transfer coefficient, i.e.
convection heat transfer is only considered to the extent
that it provides a boundary condition for the conduction problem.
The heat flux is proportional to the temperature difference
between the surface and the far field \cite{Incropera} i.e.
\[
k ~ \p T/ \p n = \htc~ (T_{\infty}-T_{\mbox{{\scriptsize {\it surface}}}}).
\]
Isothermal conditions can be realized in the limit of
strong convection, i.e. large Biot number.
Here, we should point out that isothermal conditions have been widely
used in the heat transfer
analysis of extended surfaces due to the very little information
available on the coupling between fin conduction and fluid convection,
and the weak dependence of the heat transfer coefficient
on the temperature difference between the base of the fin and its
tip (\cite{Karagiozis94}; \cite{Rohsenow}, 4.36; \cite{Kraus}, 4.5).

As we have mentioned, the optimization problem is an inverse design
problem in the sense that the objective is to find the
geometry that maximizes the heat transfer rate as oppose to
the classical/direct problem, where the objective is to find the heat
transfer rate associated with a given geometry. The
objective function is the Shape Factor \cite{Incropera,Hahne}, i.e. the
total heat transfer rate, and the variable of the optimization is the
shape of the pipe which is parameterized though the parameters
of the generalized Schwarz-Christoffel transformation. Hence, using
Geometry Parametrization \cite{Parte2010,Iaccarino2011,Wei2009,Allison2005},
the Shape Optimization problem is posed as a nonlinear
programming problem (constrained nonlinear optimization \cite{Fletcher}),
which is solved numerically \cite{Schittkowski}.

For regular, symmetric,
isothermal, doubly-periodic walls, the
heat conduction problem in a semi-infinite domain was addressed by
Fyrillas \& Pozrikidis \cite{Fyrillas01}
using both boundary integral and asymptotic methods.
For 2D periodic channels/slabs, the problem has been addressed
by Brady \& Pozrikidis \cite{Brady93} where the authors considered
the heat conduction problem associated with irregular
isothermal periodic surfaces, using the generalized Schwarz-Christoffel
transformation developed by Davis \cite{Davis79},
Floryan \cite{Floryan86}
and Floryan \& Zemach \cite{Floryan93}.  It was concluded that
for regular geometries the shape plays a more important role in the
total transport rate rather than the total arc-length while, for
self-similar irregularities, the height of the roughness is the significant
factor. These conclusions lead naturally to the following
question: Given the arc-length and the period of a periodic surface/curve,
what is the geometry that maximizes the overall transport rate;
This Shape Optimization problem was addressed
by Leontiou, Kotsonis \& Fyrillas \cite{Fyrillas13}.

From an engineering perspective, knowledge of a surface geometry that maximizes
the transport rate offers opportunities for new designs that exhibit enhanced
characteristics and properties. For example, the problem of transport
across an uneven surface, described in the preceding paragraph, is relevant
to a variety of engineering applications involving heat transfer
across rough and irregular boundaries, such as the surface
of a circuit board in microelectronics
\cite{Peterson90,Ankireddi08,Stavrou05,Patel03}.
In general, heat transfer in slab-like configurations
is of interest to problems associated with
Heat Transport from Extended Surfaces (Fins) \cite{Kraus}
and inverted high
conductivity fins/inserts \cite{Mazloomi12}.
Published work in
these thematic areas
\cite{Ganzelves97,Bobaru04,Moharana08,Kundu05} suggests that
there is potential
for significant improvements if one considers a two-dimensional (2-D)
heat conduction model as suggested by Aziz \cite{Aziz92}.

In Section \S2 we address the heat transfer problem associated with
a periodic array of periodic extensions/fins of
uniform convection heat transfer coefficient.
In Section \S3 we address the shape optimization
problems associated with the optimum shape of the fins such
that the heat transfer rate is maximized.
In particular, in Section \S3.2 we investigate under what
conditions a rectangular fin enhances the heat transfer rate.

\section{Shape factor of a periodic array of extended surfaces (fins)}

In this section we consider 2-D heat conduction in a finite slab.
The geometry of the slab is periodic in the horizontal direction
and bounded in the vertical direction by an isothermal ($T_0$) flat
surface at the top, while the bottom surface is subjected to convection with
a uniform convection heat transfer coefficient ($\htc$)
and a constant far-field temperature $T_{\infty}$ \cite{Incropera}.
The bottom surface is not flat, rather a periodic array of
extensions is present (extended surfaces, fins), in order to enhance
the heat transfer rate \cite{Incropera,Kraus}. Continuity of the
heat flux at the bottom surface implies that the heat conduction rate,
$k~{\bf n} \cdot \overrightarrow{\nabla} T = k~\p T / \p n$, must be equal to
the heat convection rate $\htc\,(T-T_{\infty})$, where $k$ is the thermal conductivity.

\begin{figure}[h!]
\renewcommand{\baselinestretch}{1}
\center{\includegraphics[width = 0.6\textwidth, clip = true]{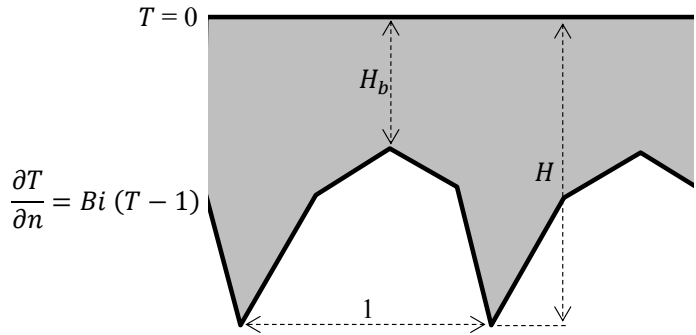}}
    \caption{Schematic representation of the problem in the physical domain.
    All variables are non-dimensional; lengths are non-dimensionalized with the
    distance between the periodic fins, i.e. period $L$. The dimensionless thickness
    of the base of the fin is $H_b$, and the length of the fin is $H_f=H-H_b$.
    The non-dimensional temperatures are $T=0$ at the top boundary and $T_{\infty}=1$
    at the far field. }
    \label{Figure:physical_domain}
\end{figure}

We non-dimensionalize lengths with the distance between
the fins (period $L$), and the temperature by subtracting $T_0$
and dividing by the temperature difference $T_{\infty}-T_0$. The dimensional
analysis leads to the following definition for the Biot number,
$Bi=L~\htc/k$.  The domain and the dimensionless parameters associated
with the problem are clearly indicated in Fig. \ref{Figure:physical_domain}.

\subsection{Conformal transformation of the physical domain into
a straight channel}

To find the heat transfer rate of such a periodic slab we first
transform it into a straight channel
\cite{Fyrillas09,Fyrillas08,Fyrillas10,Fyrillas13,Fyrillas14,Fyrillas15}.
The relevant transformation, the generalized Schwarz-Christoffel
transformation applicable for periodic channels,
has been developed by Davis and Floryan \cite{Davis79,Floryan86}
\begin{equation}
z(\hat{w};\boldsymbol{\alpha}) = R~\int_0^{\hat{w}} ~
\prod_{l=-\infty}^{l=\infty} \prod_{j=0}^{j=N}
\left( \sinh \left[\frac{w_N\,\pi}{2\,h}
 (\hat{\theta} - \hat{w}_j -l)\right] \right)^{\alpha_j} ~\measd \hat{\theta},
 \label{integral SC bounded}
\end{equation}
where the inner product identifies the number of elements ($N$) of the
discretized lower boundary (fin), and the infinite outer
product the periodic nature of the domain. We normalize lengths in the
complex domain with $w_{\sN}$, i.e. the period in the
transformed domain. Hence, in the above transformation, $\hat{w}_j$s are the
normalized images of the $z_j$s vertices, $\boldsymbol{\alpha}$ represents the
$N$+1-tuple $\alpha_0, \alpha_1, \alpha_2, \ldots, \alpha_{\sN}$ which
are equal to the turning angles multiplied by $\pi$ (the angles are taken to
be positive for a clockwise rotation, and $\alpha_0$ and $\alpha_{\sN}$ are
defined with respect to the $x$-axis),
$R$ is a complex constant, and $\hat{h}$ is the normalized height of the
channel in the  transformed domain (without loss of generality we assume that
$h=H$). For the configurations we will consider $R$ is a real number and
can be obtained by requiring that the upper wall of the physical plane,
i.e. the line $z=\mbox{i}\,H$, transforms to $w=\mbox{i}\,h$:
\begin{equation}
\Im \left[z[\mbox{i}\,\hat{h};\boldsymbol{\alpha}] \right]=H.
\label{Req}
\end{equation}
In addition, in view of the geometry, we must have
\begin{equation}
\sum_{j=0}^{N} \alpha_j=0.
\label{sumalpha}
\end{equation}

Given the domain, the parameters of the transformation
(\ref{integral SC bounded}) can  be calculated by
solving a system of non-linear equations \cite{Davis79,Floryan86}.
However, in this work, we pose an optimization problem
where the lower boundary is the variable of the optimization.
In particular, we look for the
optimum shape of the extended surface such that the heat transport rate
is maximized. We pose the problem with respect to the parameters
$\boldsymbol{\alpha}$'s of the Schwarz-Christoffel transformation.
Essentially, we parameterize the lower boundary
\cite{Parte2010,Iaccarino2011,Wei2009,Allison2005}
 with respect to
the parameters $\boldsymbol{\alpha}$'s, which are the variables
of the optimization procedure, while the objective function is
the heat transfer rate (the Shape Factor). An expression for the
Shape Factor is obtained in the following section using the
boundary element method.
\begin{figure}[h!]
\center{ \includegraphics[clip=true]{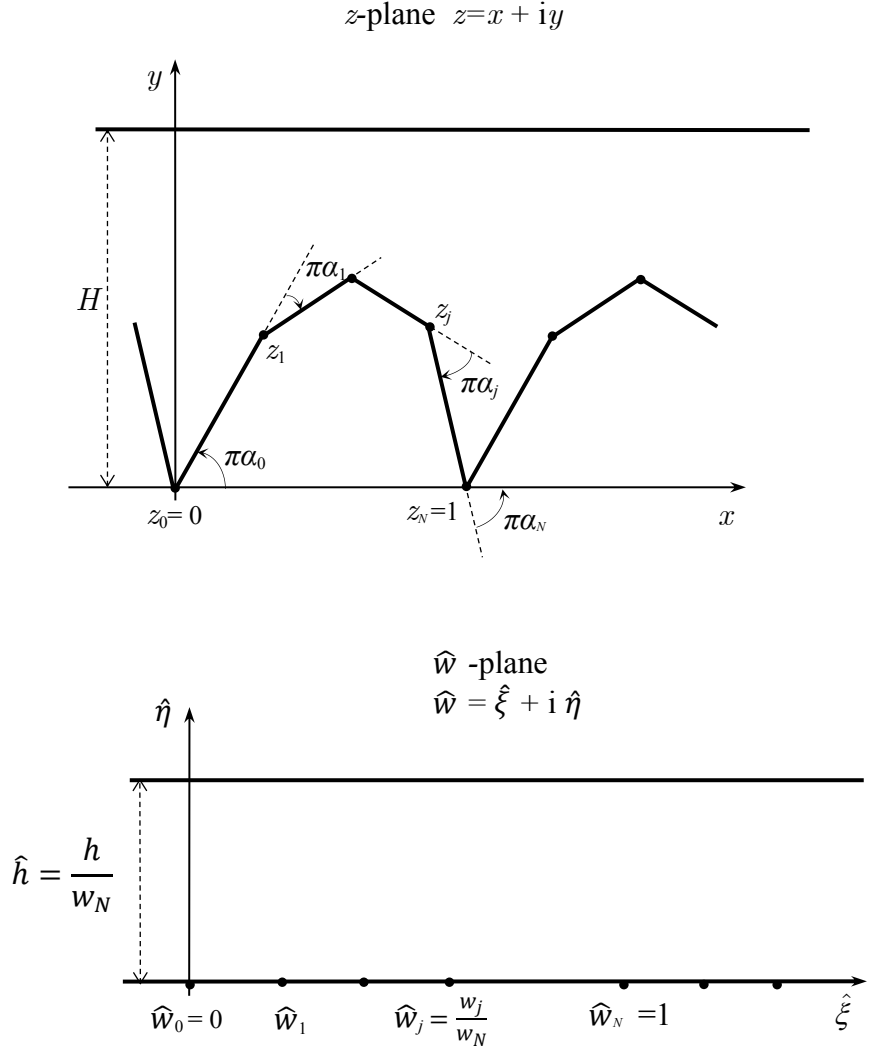}}

\caption{Mapping of the physical domain onto a straight channel using the
    generalized Schwarz-Christoffel transformation
    (equation \ref{integral SC bounded}). Lengths in the physical and
    complex domain are non-dimensionalized with $L$ and $w_{\sN}$,
    respectively.}
    \label{Figure:physical_to_complex.eps}
\end{figure}

\subsection{Shape Factor of an extended surface (fin)}

In view of the conformal transformation and the boundary condition
on the lower surface we can obtain the following expressions for
the Shape Factor ($S$) \cite{Incropera} associated with an extended
surface of unit span:
\begin{equation}
S= -\int_0^1 \frac{\p T}{\p \hat{\eta}} [\hat{\xi},\hat{\eta}=0]~\measd \hat{\xi} =
 Bi\,\int_0^1 \left(1-T[\hat{\xi},\hat{\eta}=0] \right)
\left\lvert \frac{\displaystyle \measd z}{\displaystyle \measd \hat{w}}
\right\rvert_{\hat{w}=\hat{\xi}} ~\measd \hat{\xi}.
\label{Shape Factor}
\end{equation}
We also define the fin effectiveness ($\varepsilon_f$) as the ratio of the
Shape Factor, as defined above, to the Shape Factor of the base
without fins, i.e. $ S_b=Bi/(1+Bi~H_b) $:
\begin{equation}
\varepsilon_f=\frac{S}{S_b}.
\label{effectiveness}
\end{equation}
Hence, the definition of the Shape Factor and the fin effectiveness
associated with a fin
departs from the classical definitions \cite{Incropera}, as it includes
the area of the base not covered by the fin.
{\it Based on the above definition, an addition of an extended surface
or fin would improve the heat transfer rate of the base, if
its effectiveness is greater than one.}

The temperature along the lower surface, which includes the fin,
can be obtained by applying the boundary element method
\cite{Brebbia,Stone,Fyrillas00,Fyrillas04,Ioannou12,Fyrillas01,Fyrillas11}.
An appropriate Green's function is that associated with
a periodic array of sources of period $1$ located along an insulated
lower surface and a Dirichlet boundary condition along the top surface
as described in \cite{Fyrillas08,Fyrillas10,Fyrillas14,Fyrillas15}:
\begin{equation}
G[\hat{\xi}'-\hat{\xi},\hat{\eta}'=0]= -\left(\hat{h} + \frac{1}{\pi}
\displaystyle\sum_{m=1}^{\infty} \frac{1}{m} \tanh\left[2\,\pi\,m\,\hat{h}\right]
\cos\left[2\,\pi\,m\,(\hat{\xi}'-\hat{\xi})\right] \right).
\label{GreensFunction}
\end{equation}
\begin{figure}[h!]
\center{\includegraphics[clip=true]{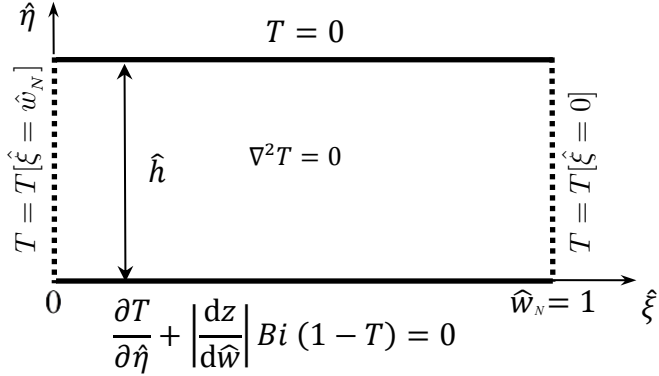}}
\begin{picture}(0,0)(0,0)
   \setbox0=\hbox{{\mbox{ $T=T[\hat{\xi}=\hat{w}_{\sN}]$}}}
   \put (-265,-30) {\shortstack[l] {\rota0{-90}}}
   \setbox0=\hbox{{\mbox{ $T=T[\hat{\xi}=0]$}}}
   \put (-55,-20) {\shortstack[l] {\rota0{-90}}}
\end{picture}

\caption{Schematic representation of the model problem along with
    boundary conditions. The lower boundary, which represents the extended
    surface, is subjected to convection heat transfer. Note that the conformal
    transformation has introduced an artificial heat transfer coefficient
    that reflects the transformation of the geometry.}
\label{Figure:domain_mathematical}
\end{figure}

The mathematical domain along with the boundary conditions is shown in
Fig. \ref{Figure:domain_mathematical}.
Applying the boundary element method and the boundary conditions we obtain
a Fredholm Integral equations of the second kind for the
temperature along the lower surface,
\begin{equation}
T[\hat{\xi}]=Bi\,\int_0^1 G\left[\hat{\xi}'-\hat{\xi}\right]
\left\lvert \frac{\displaystyle \measd z}{\displaystyle \measd \hat{w}}
\right\rvert_{\hat{w}=\hat{\xi}} \left(T[\hat{\xi}]-1 \right) ~\measd \hat{\xi}.
\label{BLE}
\end{equation}

The integral equation is solved numerically using the collocation boundary element
method \cite{Brebbia,Stone,Fyrillas00,Fyrillas04,Fyrillas08,Fyrillas10,Fyrillas11,
Ioannou12,Fyrillas14}. It is important to note that in the above
formulation the Generalized Schwarz-Christoffel transformation appears in the
boundary element formulation similar to \cite{Fyrillas15}. Another complication
is that unlike \cite{Fyrillas15}, the singularities of the transformation also
appear in the boundary element formulation, which are addressed using
Gauss-Jacobi quadrature as explained in the following section, where
we pose the Shape Optimization problem. The objective function is the
fin effectiveness ($\varepsilon_f$), i.e. the normalized Shape
Factor, and the variable of the optimization is the shape of the lower surface
(fin); the latter is characterized through the parameters of the generalized
Schwarz-Christoffel transformation.

\section{Optimum extended surfaces} \label{Optimization}

The formulation of the Shape Optimization problem follows along the same
lines as the problems
formulated by Fyrillas and Leontiou \& Fyrillas
\cite{Fyrillas09,Fyrillas08,Fyrillas10,Fyrillas13,Fyrillas14,Fyrillas15}.
The objective function is the fin effectiveness (Eq. \ref{effectiveness}) and
the constraints are dictated in view of the geometrical configuration
(Fig. \ref{Figure:physical_to_complex.eps}):
\[
\begin{array}{cr}
\hspace{0.1in} \mbox{\bf{maximize}} \hspace{0.1in} &
\varepsilon_f[\boldsymbol{\alpha},\hat{w}{\sN}]  \\[-0.0in]
\left(\boldsymbol{\alpha},\hat{w}{\sN} \right) &
\end{array}
\]
subject to the constraints
\begin{equation}
\begin{array}{l}
  \sum_{j=0}^{N-1} |z_{j+1}-z_j|= P \\[0.1in]
  \sum_{j=0}^{N} \alpha_j=0 \\[0.1in]
  \Im[z_i]= y_i \geq (H-H_b) \\[0.1in]
  \Re[z_{\sN}]=x_{\sN}=1 \\[0.1in]
  \Im[z_{\sN}]=y_{\sN}=0,
\end{array}
\label{MaxS}
\end{equation}

where the perimeter $P$, the height $H$ and the height
of the base $H_b$ are assigned a priori.
Note that the period is equal to one as it is used for
non-dimensionalization, and the third constraint is an explicit
equation to obtain the real constant $R$ (Eq. \ref{Req}).
The fourth constraint defines the length of the
fin and hence, the thickness of the plate.
The equality $H_f=H-H_b$ is achieved for sufficiently
large perimeter $P$.

Similar to our previous work
\cite{Fyrillas09,Fyrillas08,Fyrillas10,Fyrillas13,Fyrillas14,Fyrillas15}
the integral $\hat{z}_i$  (Eq. \ref{integral SC bounded}) is calculated using Gauss-Jacobi
quadrature \cite{Driscoll}, and we choose the $\hat{w}_j$s to be
equispaced between 0 and 1. The infinite product appearing in integral
(\ref{integral SC bounded}) can be truncated to a
small value without affecting the accuracy due to
the exponential decay of the hyperbolic sines \cite{Floryan86}.
The numerical optimization has been performed using
the NLPQL optimization code developed by Schittkowski \cite{Schittkowski}.

\begin{figure}[h!]
\renewcommand{\baselinestretch}{1}
\begin{minipage}[c]{.9\linewidth}
\centering\epsfig{figure=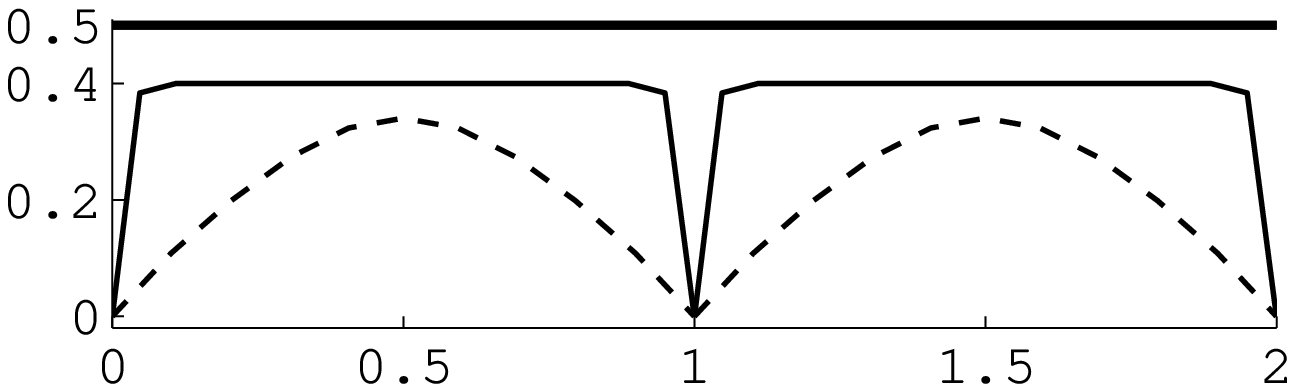,width=0.7\linewidth,clip=true}
\end{minipage}

\vspace*{-1.3in}

\begin{minipage}[c]{.9\linewidth}
\centering\epsfig{figure=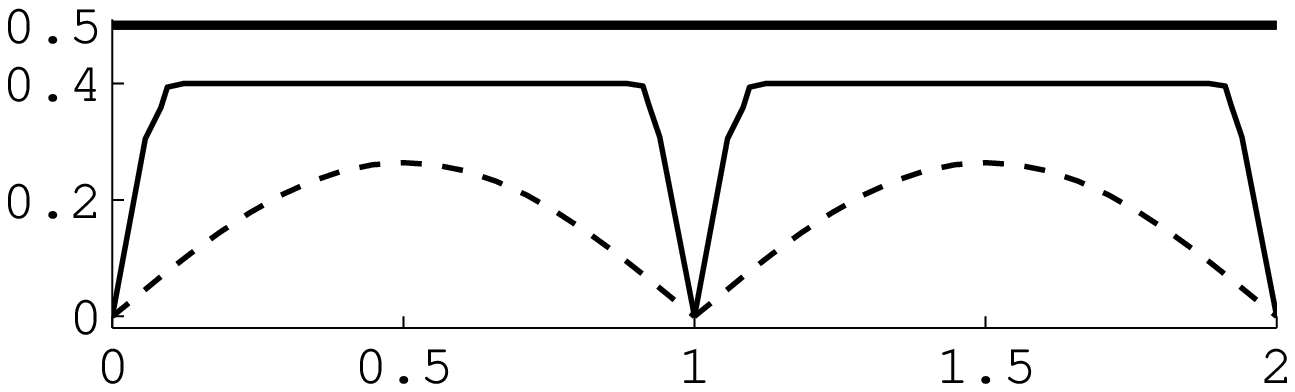,width=0.7\linewidth,clip=true}
\end{minipage}

\vspace*{-1.3in}

\begin{minipage}[c]{.9\linewidth}
\centering\epsfig{figure=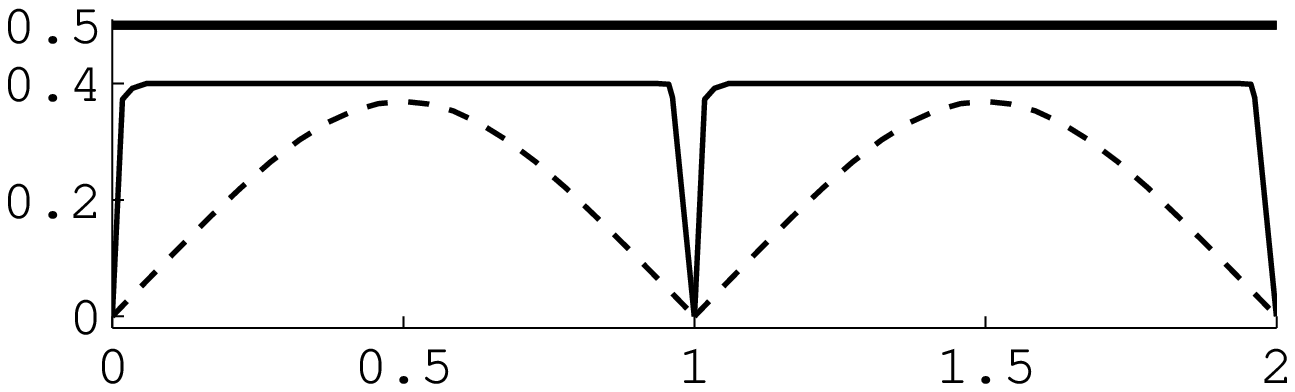,width=0.7\linewidth,clip=true}
\end{minipage}

\begin{picture}(0,0)(0,0)
   \put (100,500)  {\shortstack[l]  (a) $Bi=1$}
   \put (65,460)   {\shortstack[l]  {$y$}}
   \put (200,410)  {\shortstack[l]  {$x$}}

   \put (100,335)  {\shortstack[l]  (b) $Bi=10$}
   \put (65,300)   {\shortstack[l]  {$y$}}
   \put (200,245)  {\shortstack[l]  {$x$}}

   \put (100,175)  {\shortstack[l]  (c) $Bi=100$}
   \put (65,130)   {\shortstack[l]  {$y$}}
   \put (200,85)  {\shortstack[l]   {$x$}}

\end{picture}

\vspace*{-1.2in}

\caption{Optimum surfaces/fins that maximize the heat
transfer rate in a finite channel of height $H=0.5$. The fins are assumed
to be attached to a base whose minimum thickness is
$H_f=H-H_b=0.1$. In {\bf Fig. (a)} we show results for $Bi=1$ and
two different perimeters: $P=1.24$ (dashed curve, $\varepsilon_f=1.0$);
$P=1.68$ (solid curve, $\varepsilon_f=1.4$).
In {\bf Fig. (b)} we show results for $Bi=10$ and
two different perimeters: $P=1.15$ (dashed curve, $\varepsilon_f=0.5$);
$P=1.63$ (solid curve, $\varepsilon_f=1.0$).
In {\bf Fig. (c)} we show results for $Bi=100$ and
two different perimeters: $P=1.27$ (dashed curve, $\varepsilon_f=0.5$);
$P=1.72$ (solid curve, $\varepsilon_f=0.99$).}
\label{Fig:MaxwN}
\end{figure}

In Fig. \ref{Fig:MaxwN} we show numerical results of the fin
effectiveness maximization problem (Eq. \ref{MaxS}) associated
with a slab of height $H=0.5$ and $H_b=0.1$,
for different values of the perimeter ($P$) and Biot number ($Bi$).
The results suggest that in all cases the optimum fin is infinitely thin, elongated in the
vertical direction. It is very important to note that for
a large Biot number the presence of the fin might not enhance
the heat transfer rate from the base, i.e. the
addition of a fin reduces the heat transfer rate. This can be
justified as in the case of isothermal conditions the addition
of a fin would result in a reduction of heat transfer
(see Introduction, \S1 first paragraph).

The above results/conclusions suggest that:
(i) the optimum fin is infinitely thin and elongated, and
(ii) there exists a critical Biot number which characterizes whether
a fin is effective or not.
Hence, in the next section we determine the critical Biot
number of a rectangular fin.

\subsection{Critical thickness characterizing the effectiveness of a fin}

The results obtained through the Shape Optimization procedure,
suggest that for a fixed perimeter the optimum fin is an
infinitely thin extended surface. It is tempting to infer that this is an
artifact of the uniform heat transfer coefficient; had
conjugate heat transfer been considered, the optimization procedure
would have led to a different result. Furthermore, an
interesting result that has emerged from the optimization analysis is that
there exists a critical Biot number, associated with a given fin geometry,
that determines whether the
addition of the fin would enhance the heat transfer rate.
In this section, we elucidate this point by considering
the heat transfer rate associated with particular geometries,
i.e. we consider the effectiveness of rectangular fin \cite{Incropera}.

\begin{figure}[h!]
\begin{minipage}[c]{.5\linewidth}
\centering\epsfig{figure=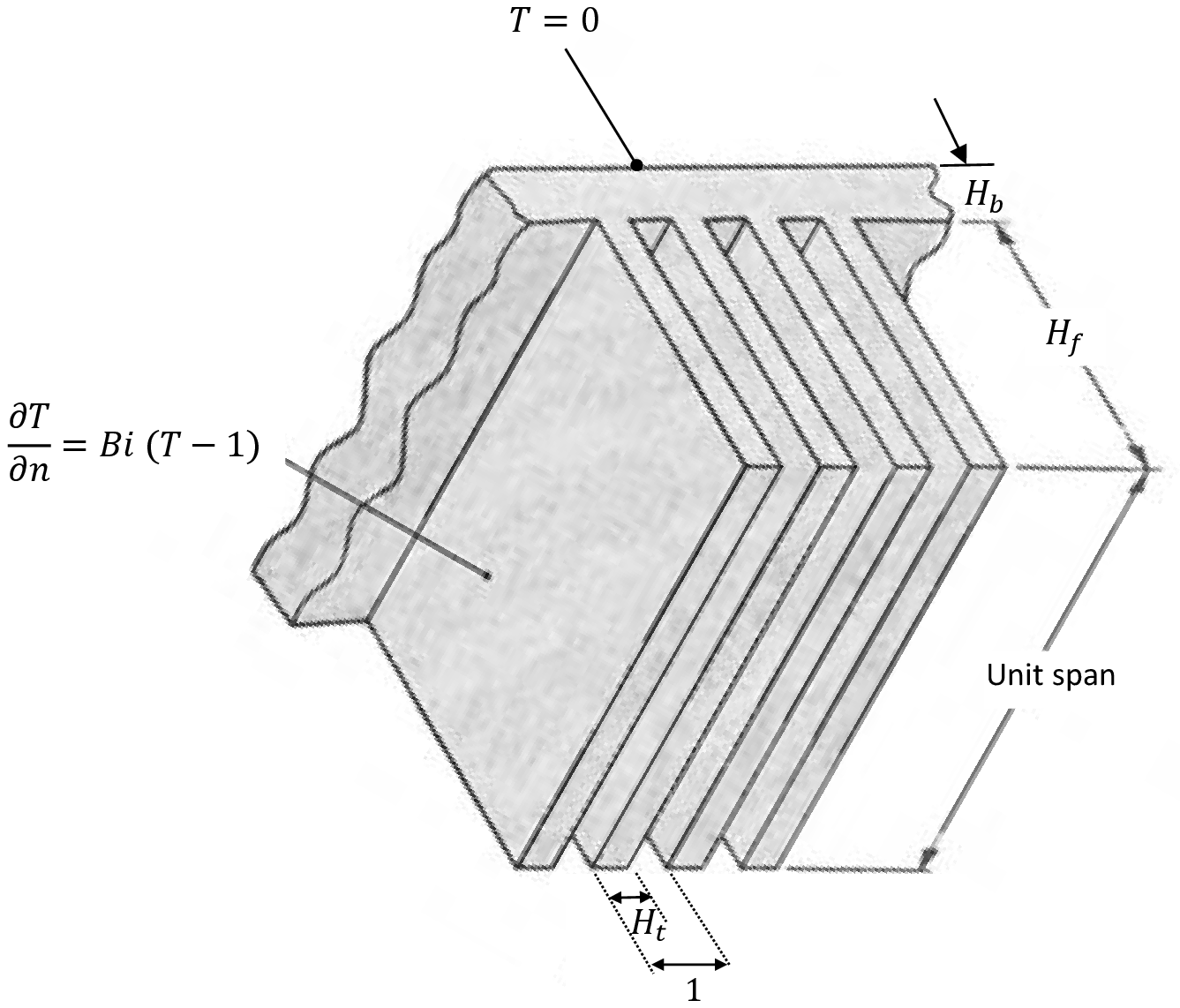,width=1\linewidth,clip=true}
\end{minipage}
\begin{minipage}[r]{.5\linewidth}
\centering\epsfig{figure=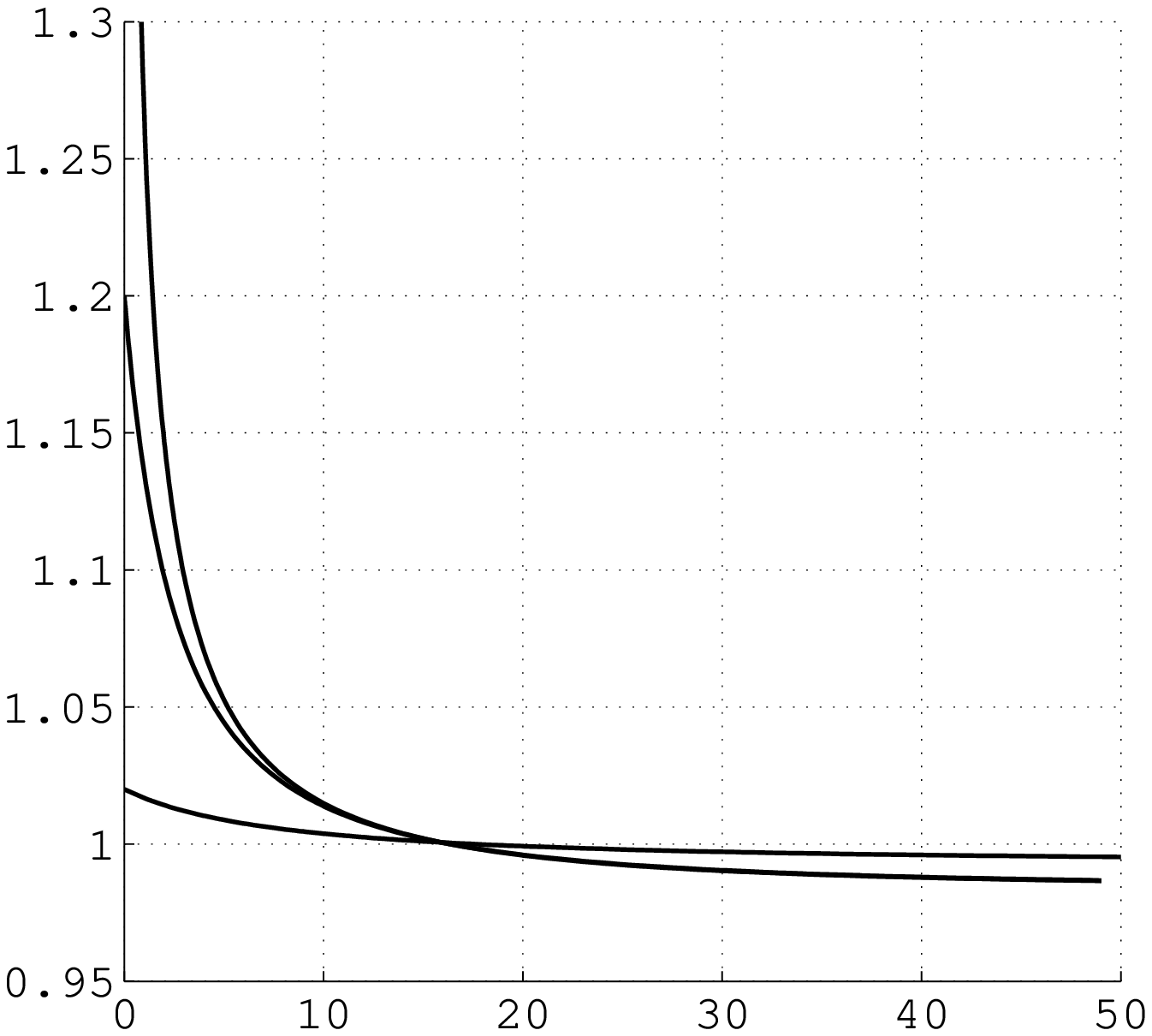,width=0.7\linewidth,clip=true}
\end{minipage}

\begin{picture}(0,0)(0,0)
   \put (50,170)  {\shortstack[l]  (a)}

   \put (270,170)  {\shortstack[l]  (b)}
   \put (250,100)   {\shortstack[l]  {$\varepsilon_f$}}
   \put (330,20)  {\shortstack[l]  {$Bi$}}
\end{picture}

\caption{Fig (a): A periodic array of rectangular fins.
Fig (b): Fin effectiveness ($\varepsilon_f$) Vs Biot number ($Bi$)
for a rectangular fin of $H_b=H_t=0.1$. The three curves
correspond to three different lengths ($H_f$).
The lower curve corresponds to $H_f=0.01$ while the middle
and top curves to $H_f=0.1$ and $H_f=1$, respectively.}
\label{Fig:rectangular_fin}
\end{figure}

We consider the classical configuration of a periodic array
of rectangular fins attached to an infinite rectangular base
Fig \ref{Fig:rectangular_fin}a. Similar to the previous section,
we non-dimensionalize lengths with the period of the array
($L$), to obtain the dimensionless Biot number $Bi=\htc L/K$.
In Fig. \ref{Fig:rectangular_fin}b, we show a plot of the
fin effectiveness Vs the Biot number for a rectangular fin of
thickness $H_t=0.1$, attached to base of thickness $H_b=0.1$.
The critical Biot number, i.e. the value of the Biot number
where the effectiveness of the fin is equal to one, is
$Bi_{critical} \approx 16.4$ and it is independent of the
length of the fin $H_f$. We have considered different
sizes of rectangular fins and concluded that:
\begin{enumerate}
\item {\bf A rectangular fin is effective if $Bi~H_t \leq 1.64$.
Expressed in dimensional variables, {\it fin thickness}} $< 1.64~k/\htc$.
This is independent of both the thickness of the base and the length of
the fin.
\item \bf{ The maximum effectiveness is realized at $Bi=0$ and is equal to}
$\varepsilon_f \left[ Bi=0 \right]= 2 H_f + 1$.
\end{enumerate}
The results have
been verified through finite difference and finite element simulations.

\section{Conclusions}

In this work we consider the heat transfer, shape optimization
problem associated with a periodic array of extended surfaces
subjected to  convection with a uniform heat transfer
coefficient. We address both the heat transfer problem and
the Shape Optimization, inverse design problem, of
finding the geometry that maximizes the heat transfer rate.

The problem is formulated as a two dimensional, arbitrary channel
of unit length which is bounded from above by a flat isothermal surface
and from below by periodic extensions/fins with a uniform
heat transfer coefficient.
Given a set of geometrical constraints that characterize the
geometry of the fin and the base, the objective is to
find the optimum shape of the fin such that the heat transfer rate
is maximized. Within the approximation of uniform heat transfer
coefficient, the optimization procedure suggests
that the optimum fin is infinitely thin and long.


Furthermore, the optimization procedure has revealed a very
interesting result. There is a critical Biot number that
characterizes the fin effectiveness. For values of the
Biot number less than the critical a fin enhances
the heat transfer rate, while for higher values it
attenuates the heat transfer rate, hence it is not effective.
This is investigated further by considering the simplest case
of a rectangular fin of uniform cross section. Numerical results
have elucidated the fin effectiveness and are summarized as follows:
\begin{enumerate}
\item {\bf A rectangular fin is effective if $Bi~H_t \leq 1.64$.
Expressed in dimensional variables, {\it fin thickness}}
$< 1.64~k/\htc$.
This is independent of both the thickness of the base and the length of
the fin.
\item {\bf The maximum effectiveness is realized at $Bi=0$ and is equal to}
$\varepsilon_f \left[ Bi=0 \right]= 2 H_f + 1$.
\end{enumerate}

{\bf Acknowledgment.}\\
The work was funded by Porfyrios Chap Glass Ltd. The authors would like to thank
Klaus Schittkowski for providing his NLPQL numerical optimization code \cite{Schittkowski}.




\end {document}